\documentclass[twocolumn,pre,preprintnumbers,amsmath,amssymb]{revtex4-1}

\usepackage{color}
\definecolor{darkgreen}{rgb}{0,0.5,0}
\definecolor{blue}{rgb}{0,0,0.8}
\definecolor{lightblue}{rgb}{0.93,0.96,1}
\definecolor{darkblue}{rgb}{0.,0.,0.6}
\usepackage[colorlinks,linkcolor=darkgreen,citecolor=darkblue,urlcolor=blue]{hyperref}

\usepackage{graphicx}
\usepackage{dcolumn}
\usepackage{bm}
\usepackage{multirow}
\usepackage{natbib}
\usepackage{arydshln,amsmath,amssymb,float}

\def\reff#1{(\ref{#1})}

\renewcommand\matrix[1]{\,\mathrm{\underline{#1}}\,}

\usepackage{lineno}

\DeclareGraphicsExtensions{.png,.gif,.jpg,.pdf,.bmp}

\begin{document}

\title{Rigorous elimination of fast stochastic variables from the linear noise \\ approximation using projection operators}


\author{Philipp Thomas}
\affiliation{Department of Physics, Humboldt University of Berlin, Newtonstr. 15, D-12489 Berlin, Germany}
\affiliation{School of Biological Sciences, University of Edinburgh, Edinburgh EH9 3JR, United Kingdom}

\author{Ramon Grima}
\affiliation{School of Biological Sciences, University of Edinburgh, Edinburgh EH9 3JR, United Kingdom}

\author{Arthur V. Straube}
\affiliation{Department of Physics, Humboldt University of Berlin, Newtonstr. 15, D-12489 Berlin, Germany}

\begin{abstract}

The linear noise approximation (LNA) offers a simple means by which one can study intrinsic noise {in} monostable biochemical networks. Using simple physical arguments, we have recently introduced
the slow-scale LNA (ssLNA) which is a reduced version of the LNA under conditions of timescale separation. In this paper, we present the first  {rigorous derivation} of the ssLNA using {the} projection operator technique and show that the ssLNA follows uniquely from the standard LNA under the same conditions of timescale separation as those required for the deterministic quasi-steady state approximation. We also show that the large molecule number limit of several common stochastic model reduction techniques under timescale separation conditions {constitutes} a special case of the ssLNA.
\end{abstract}

\maketitle

\section{Introduction}


In the study of complex systems, it is common to invoke assumptions under which the dimension (and hence complexity) of the system is reduced; such a strategy often leads to relatively simple theories capable of exact analytical predictions, which offer insights typically lost in numerical approaches. This approach is particularly useful in the study of biochemical pathway dynamics which typically involve the interaction of hundreds or thousands of different chemical species \cite{alon-book-07}. Deterministic models of such systems are frequently simplified by invoking the presence of well-separated timescales \cite{alon-book-07, alon-nature-07}, in particular by means of the quasi-steady-state approximation (QSSA) \cite{segel-slemrod-89, lee-othmer-09}. However it is well appreciated nowadays that a discrete stochastic approach {in terms of chemical master equations (CMEs)} is more faithful in describing dynamics inside living cells since the number of molecules of many species is in the range of few tens to few thousands \cite{grima-08}. The development of reduced stochastic models consistently coarse-grained over timescales presents a {significantly} larger challenge than the reduction of deterministic models because of the much larger number of differential equations which need to be solved in the former compared to the latter.

\begin{figure}
\includegraphics[scale=0.65]{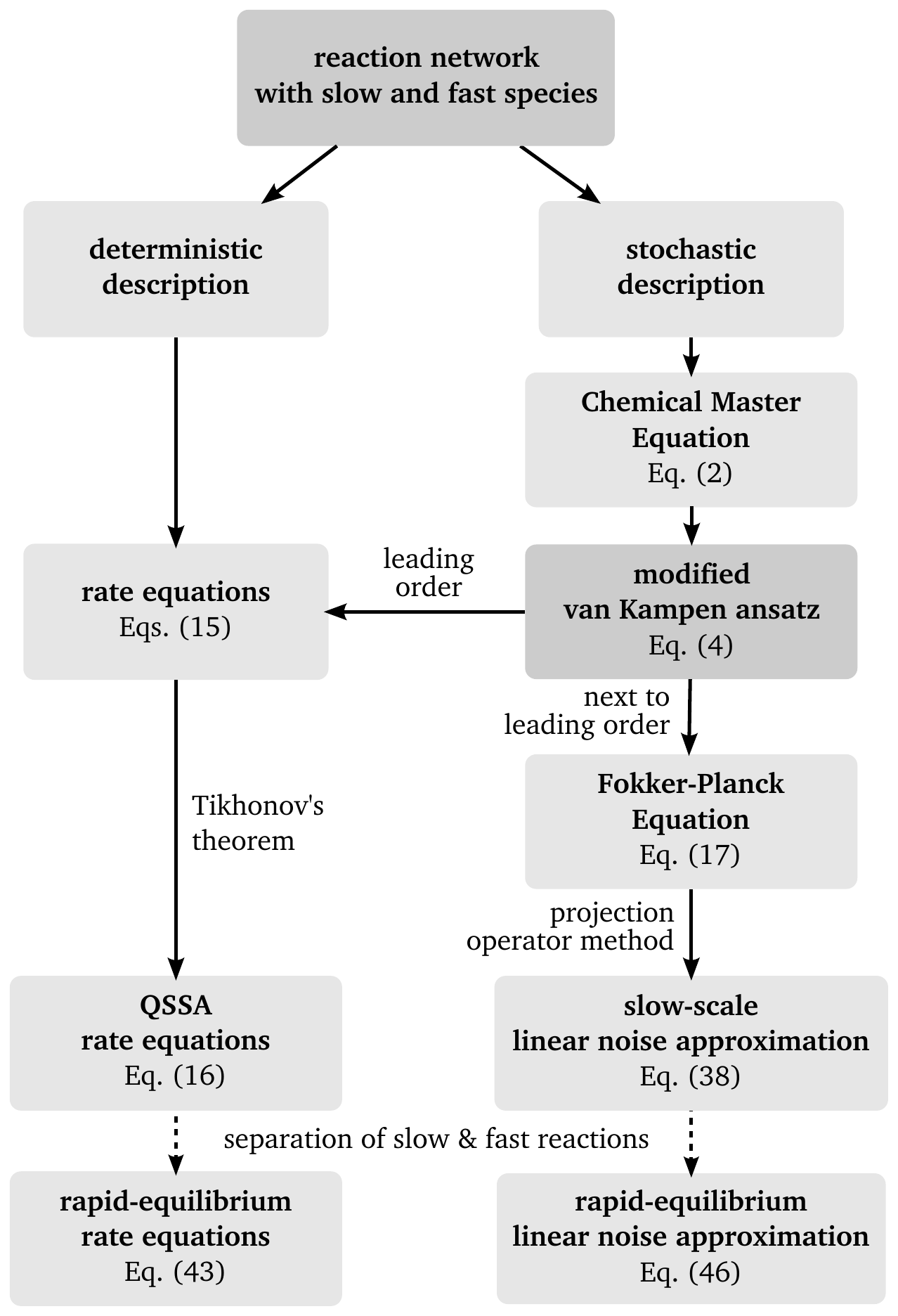}
\caption{Schematic {illustrating} the derivation of deterministic and
stochastic quasi-steady state approximations and rapid-equilibrium
approximations for a reaction network with slow and fast species. The
deterministic approach based on {rate equations} uses Tikhonov's theorem while the
stochastic approach utilizes the projection operator method applied to
the system size expansion derived from a modified van Kampen ansatz.
}
\end{figure}

Indeed it has been shown that it is not possible to formulate a reduced CME for the whole region of parameter space in which the deterministic QSSA is valid but rather only over a subset of this space \cite{janssen-89,mastny-07,thomas-jcp-comm-11}. For example consider the Michaelis-Menten mechanism in which substrate reversibly binds with enzyme to form an enzyme-substrate complex which then decays into product. We are interested in conditions such that the substrate species decays over a much longer timescale than the enzyme and complex species. For this case, a consistently reduced CME can only be written provided that the complex decay rate into product is much less than its decay rate into substrate and enzyme which is not necessarily the case \cite{sanft-11,thomas-jcp-comm-11}. However it is possible to write down reduced deterministic rate equations (REs) using the QSSA even when the aforementioned condition is not satisfied! \cite{thomas-jcp-comm-11}

Recently we have shown that a reduced stochastic description with the same range of validity as the deterministic QSSA is possible. This reduction is achieved by first approximating the CME by linear Langevin equations, an approximation called the {linear noise approximation} (LNA) and then integrating out the fast fluctuations such that one obtains a reduced set of linear Langevin equations for the fluctuations in the slowly varying species only. The latter is the slow-scale LNA (ssLNA)  \cite{thomas-bmc-12}. 
The advantage of the ssLNA over the reduced CME approach is that the former is valid over the same range of parameter space as the QSSA, a claim which has been numerically verified for a number of biochemical circuits {including cooperative and non-cooperative enzyme reactions and gene networks with or without negative feedback regulation \cite{thomas-bmc-12}}. The ssLNA, as it currently stands, has been derived by means of simple and intuitively clear physical arguments but is lacking a formal and rigorous derivation. In this article we provide such a derivation and also show the relationship of the ssLNA with other common methods of stochastic model reduction based on timescale separation assumptions. 

The paper is divided as follows. In Sec.~\ref{full-LNA}, we consider the CME of a general system of elementary chemical reactions with two characteristic and clearly separated timescales and reformulate the conventional system size expansion of the CME under such conditions. 
In Sec.~\ref{sec:ssLNA} we use the leading order terms of the expansion to show that the deterministic limit of the CME under timescale separation conditions is formally the same as the conventional reduced rate equations obtained from the QSSA. The next to leading order terms of the expansion provide us with a Fokker-Planck equation which we reduce to a simpler Fokker-Planck equation for the slow fluctuations only, by means of the projection operator formalism. The reduced Fokker-Planck equation is obtained in closed-form for all monostable reaction networks and is the same as the ssLNA obtained in \cite{thomas-bmc-12}. Finally in Sec.~\ref{sec:rel} we show how the large molecule number limit of conventional stochastic reduction methods leads to a special case of the ssLNA and that hence these methods provide results in a more restrictive range of parameter space than the ssLNA does. The methodology of our approach is summarized in Fig. 1.

\section{The system size expansion under timescale separation conditions}\label{full-LNA}

We consider a system that comprises $N$ chemical species $X_i$ ($i=1,\ldots, N$) confined in a compartment of volume $\Omega$ and assume that the species can interact via $j=1,\ldots, R$ elementary chemical reactions
\begin{equation}
\sum_{i=1}^N s_{ij} X_i  \xrightarrow{k_j} \sum_{i=1}^N r_{ij} X_i \,, \label{reaction}
\end{equation}
where $s_{ij}$ and $r_{ij}$ are the stoichiometric coefficients \cite{heinrich-schuster-book-96} and $k_j$ is the macroscopic reaction {rate}. {The constraint $\sum_i s_{ij} \le 2$ for all $j$ implies that the reactions are unimolecular or bimolecular and hence elementary.} The total number of species $N$ is assumed to comprise $N_s$ slow and $N_f=N-N_s$ fast species, respectively. For convenience, we stick to the convention that $X_1$ to $X_{N_s}$ denote the slow species, while $X_{N_s+1}$ to $X_N$ are reserved for the fast species.

In what follows, matrices are denoted by underlined quantities and all vectors are column vectors. 
Let $n_i$ denote the number of molecules of species $i$, then the probability $P(\vec{n},t)$ to find the system in a particular state $\vec{n}=(n_1,...,n_N)^T$ is determined by the {CME}  \cite{vankampen-book-07, gillespie-rev-07}
\begin{align}
&\frac{\partial P(\vec{n},t)}{\partial t} = \Omega \sum_{j=1}^{R} \biggl( \displaystyle\prod_{i=1}^N E_i^{-S_{ij}} - 1 \biggr) \hat{f}_j(\vec{n},\Omega) P(\vec{n},t)\, \label{CME}
\end{align}
where $ S_{ij}=(\matrix{S})_{ij}=r_{ij}-s_{ij}\,$ is the stoichiometric matrix and $E_i^{z}$ is a step operator, the action of which on some function of the absolute number of molecules results in the same function but with $n_i$ replaced by $n_i + z$, and $\hat{f}_j(\vec{n},\Omega)$ are the entries of the microscopic rate function vector. The product $\Omega \hat{f}_j(\vec{n},\Omega) dt$ represents the propensity function, which has the meaning of the probability that reaction $j$ takes place in a small time interval $dt$. 

The {CME} is typically analytically intractable and hence a systematic approximation method is needed. The system size expansion as developed by van Kampen is one such technique \cite{vankampen-book-07}. The heart of the method is the ansatz that the molecular concentration of the CME can be written as: 
\begin{align}
\frac{\vec{n}}{\Omega} = \vec{\phi} + \Omega^{-1/2} \vec{\eta},
\label{VKoriginalansatz}
\end{align}
where $\vec{\phi}$ is vector of concentrations as given by the corresponding {REs} and the second term on the right hand side is a stochastic term describing fluctuations about the concentrations of the REs.

We now impose timescale separation conditions, i.e., we assume that the transients in concentrations of a group of species decays much slower than those of the remaining group of species. The first group of species we label as the slow species while the second are the fast species. The characteristic timescales of these two are $\tau_s$ and $\tau_f${,} respectively. The vectors of molecular populations, of macroscopic concentrations and of fluctuations can be divided into subpopulations of slow and fast species: $\vec{n}=(\vec{n}_s,\vec{n}_f)$, $\vec{\phi}=(\vec{\phi}_s,\vec{\phi}_f)=(\tau_s \vec{x}_s,\tau_f \vec{x}_f)$ and $\vec{\eta}=(\vec{\eta}_s,\vec{\eta}_f)=(\tau_s^{1/2} \vec{\epsilon}_s,\tau_f^{1/2} \vec{\epsilon}_f)$. The ansatz Eq. (\ref{VKoriginalansatz}) can then be written as:
\begin{align}
\label{LNA-ansatz}
\begin{split}
&\frac{1}{\tau_s}\frac{\vec{n}_s}{\Omega}=\vec{x}_s+(\Omega\tau_s)^{-1/2}\vec{\epsilon_s}\,,
\\
&\frac{1}{\tau_f}\frac{\vec{n}_f}{\Omega}=\vec{x}_f+(\Omega\tau_f)^{-1/2}\vec{\epsilon_f}\,.
\end{split}
\end{align}
%
We now use this ansatz to derive an expansion of the CME in powers of the small parameter $\Omega^{-1/2}$ valid in the case of large volumes or, equivalently, for large copy numbers of molecules. For convenience the components of the various vectors will be denoted as follows: $\vec{x}_{s}=(x_1,...,x_{N_s})^T$, $\vec{x}_{f}=(x_{N_s+1},...,x_N)^T$, $\vec{\epsilon}_s = ({\epsilon_1,...,\epsilon_{N_s}})^T$ and $\vec{\epsilon}_f = ({\epsilon_{N_s+1},...,\epsilon_{N}})^T$.
The probability distribution of molecular populations $P(\vec{n},t)$ is hence forth changed into the distribution of slow and fast fluctuations $\Pi(\vec{\epsilon}_s,\vec{\epsilon}_f,t)$. In what follows we consider how the time derivative, the step operator and the microscopic rate function vector which compose the CME Eq. (\ref{CME}) transform under the proposed ansatz Eq. (\ref{LNA-ansatz}).

\subsubsection*{Transformation of the time derivative}

Using the change of variable theorem the time derivative can be written as
\begin{align}
 \frac{\partial }{\partial t} &P(\vec{n},t) = \notag \\  &\left(\frac{\partial }{\partial t} 
+  \nabla_{s}^T  \left.\frac{d\vec{\epsilon}_s}{dt} \right|_{\vec{n}}  +  \nabla_f^T  \left. \frac{d\vec{\epsilon}_f}{dt} \right|_{\vec{n}}\right) \Pi(\vec{\epsilon}_s,\vec{\epsilon}_f,t),
\end{align}
where $\nabla_{s} = (\partial / \partial \epsilon_1,...,\partial / \partial \epsilon_{N_s})^T$ and $\nabla_{f} = (\partial / \partial \epsilon_{N_s+1},...,\partial / \partial \epsilon_{N})^T$. The time derivatives at constant $\vec{n}$, denoted by $\cdot|_{\vec{n}}$, can be computed from the ansatz Eq. (\ref{LNA-ansatz}), leading to
\begin{align}
 \frac{\partial }{\partial t} &P(\vec{n},t) = \notag \\  &\left(\frac{\partial }{\partial t} 
- \sqrt{\Omega \tau_s}  \nabla_{s}^T  \frac{d\vec{x}_s}{dt}  - \sqrt{\Omega \tau_f} \nabla_f^T  \frac{d\vec{x}_f}{dt}\right) \Pi(\vec{\epsilon}_s,\vec{\epsilon}_f,t).
\label{transtimed}
\end{align}

\subsubsection*{Transformation of the step operator}

By the definition of the step operator we have
\begin{align}
\prod_{i=1}^N& E_i^{-S_{ij}} g(n_1,...,n_N) = \nonumber\\ & g(n_1-(\matrix{S}_s)_{1,j},...,n_{N_s}-(\matrix{S}_s)_{N_s,j}, \nonumber\\ & n_{N_s+1}-(\matrix{S}_f)_{1,j},...,n_N-(\matrix{S}_f)_{N_f,j}),
\end{align}
where we have partitioned the stoichiometric matrix into two parts: $(\matrix{S}_s)_{ij} = (\matrix{S})_{ij}$ for $1 \le i \le N_s$ (the stoichiometric matrix for the slow species) and $(\matrix{S}_f)_{ij} = (\matrix{S})_{i+N_s,j}$ for $1 \le i \le N_f$ (the stoichiometric matrix for the fast species). Note that $g$ denotes some general function of the molecule numbers. Applying the ansatz Eq. (\ref{LNA-ansatz}) it follows that the above equation can be written as
\begin{align}
&\prod_{i=1}^N E_i^{-S_{ij}} g(\epsilon_1,...,\epsilon_{N}) = \nonumber\\ & g(\epsilon_1-(\Omega \tau_s)^{-1/2}(\matrix{S}_s)_{1,j},...,\epsilon_{N_s}-(\Omega \tau_s)^{-1/2}(\matrix{S}_s)_{N_s,j}, \nonumber\\ & \epsilon_{N_s+1}-(\Omega \tau_f)^{-1/2}(\matrix{S}_f)_{1,j},...,\epsilon_{N}-(\Omega \tau_f)^{-1/2}(\matrix{S}_f)_{N,j}).
\end{align}
The right hand side of the above equation can be Taylor expanded from which it follows that the step operator for the {$j^\text{th}$} reaction has the following formal expansion in powers of the inverse square root of the system volume: 
\begin{align}
 \prod_{i=1}^N& E_i^{-S_{ij}}-1 = \nonumber \\& \sum_{k=1}^\infty \frac{(-1)^k \Omega^{-k/2}}{k!}\biggl[\left(\tau_s^{-1/2} \nabla_{s}^T \matrix{S}_s + \tau_f^{-1/2} \nabla_{f}^T \matrix{S}_f\right)_j \biggr]^k.
 \label{eqn:stepoperator}
\end{align}

\subsubsection*{Transformation of the microscopic rate function vector}

We consider four general types of elementary reactions depending on the order $j$ of the reaction, for which the microscopic rate functions have been rigorously derived from microscopic {considerations} \cite{gillespie-92, gillespie-09}: (i) $\hat{f}_j(\vec{n},\Omega)=k_j$, for a zeroth-order reaction by which a species is input into a compartment; (ii) $\hat{f}_j(\vec{n},\Omega)=k_j n_u \Omega^{-1}$, for a first-order unimolecular reaction describing the decay of some species $u$; (iii) $\hat{f}_j(\vec{n},\Omega)= k_j n_u (n_u-1) \Omega^{-2}$ for a second-order bimolecular reaction between two molecules of the same species $u$; (iv) $\hat{f}_j(\vec{n},\Omega)=k_j n_u n_v \Omega^{-2}$, for a second-order bimolecular reaction between two molecules of different species, $u$ and $v$. 

With each of these cases one can also associate a macroscopic rate function, $\vec{f}$, i.e., the rate of reaction as given by the corresponding rate equations. For the four cases discussed above these are: (i)$f_j(\vec{x}_s,\vec{x}_f) = k_j$; (ii) $f_j(\vec{x}_s,\vec{x}_f) = k_j \phi_u$; (iii) $f_j(\vec{x}_s,\vec{x}_f) = k_j \phi_u^2$; (iv) $f_j(\vec{x}_s,\vec{x}_f) = k_j \phi_u \phi_v$, where $\phi_i = \tau_s x_i$ if $1 \le i \le N_s$ and $\phi_i = \tau_f x_i$ if $N_s+1 \le i \le N$. 

Given the microscopic and macroscopic rate functions for the four elementary reactions, one can write the former in terms of the latter, leading to the general result
\begin{align}
 \hat{f}_j(\vec{n},\Omega)&=f_j(\vec{x}_s,\vec{x}_f)+(\Omega\tau_s)^{-1/2} \nabla^T_{\vec{x}_s} f_j(\vec{x}_s,\vec{x}_f) \vec{\epsilon}_s \notag \\
&+(\Omega\tau_f)^{-1/2} \nabla^T_{\vec{x}_f} f_j(\vec{x}_s,\vec{x}_f) \vec{\epsilon}_f  + O(\Omega^{-1}),
\label{microfnexp}
\end{align}
where $\nabla_{\vec{x}_s}$ and $\nabla_{\vec{x}_f}$ denote the vectors of derivatives with respect to the components of the vectors $\vec{x}_s$ and $\vec{x}_f${, respectively}. 

This formula can be easily verified by considering each of the four elementary reactions discussed above. For example for case of a second-order $j^{th}$ reaction involving a slow species $u$ and a fast species $v$, we have
{
\begin{align}
\hat{f}_j(n_u,n_v,\Omega) = &k_j \frac{n_u}{\Omega}\frac{n_v}{\Omega} = k_j (\tau_s \tau_f x_u x_v + (\Omega \tau_s)^{-1/2} \tau_s \times \nonumber \\& \tau_f x_v \epsilon_u + (\Omega \tau_f)^{-1/2} \tau_s \tau_f x_u \epsilon_v) + O(\Omega^{-1}), \nonumber \\=&{f}_j(x_u,x_v) + (\Omega \tau_s)^{-1/2} \frac{\partial {f}_j(x_u,x_v)}{\partial x_u} \epsilon_u + \nonumber \\ &(\Omega \tau_f)^{-1/2} \frac{\partial {f}_j(x_u,x_v)}{\partial x_v} \epsilon_v + O(\Omega^{-1}), 
\end{align}}
\noindent
where we used the definitions of the microscopic and macroscopic rate functions given above and the ansatz Eq.(\ref{LNA-ansatz}). 

\subsubsection*{The transformed CME}

Substituting Eqs. (\ref{transtimed}), (\ref{eqn:stepoperator}) and (\ref{microfnexp}) in the CME Eq. (\ref{CME}) and rescaling time by the slow timescale $\tau = t/ \tau_s$, we obtain the transformed CME:
\begin{align}  
\label{eqn:cmeexp}
&\frac{\partial }{\partial \tau}\Pi(\vec{\epsilon}_s,\vec{\epsilon}_f,\tau)\notag\\
&-\left({\Omega}{\tau_s}\right)^{1/2} \nabla^T_{s} \left(\frac{\partial \vec{x}_s}{\partial \tau} -\matrix{S}_s\vec{f}(\vec{x}_s,\vec{x}_f)\right)\Pi(\vec{\epsilon}_s,\vec{\epsilon}_f,\tau)\notag\\
  &-\left({\Omega}{\tau_f}\right)^{1/2} \nabla^T_{f}
 \left(\frac{\partial \vec{x}_f}{\partial \tau}-\gamma\matrix{S}_f\vec{f}(\vec{x}_s,\vec{x}_f)\right) 
 \Pi(\vec{\epsilon}_s,\vec{\epsilon}_f,\tau) \notag\\
&=\Omega^{0} \left(\gamma \mathcal{L}_f + \gamma^{1/2}\mathcal{L}_\text{int} + \mathcal{L}_s  \right) {\Pi(\vec{\epsilon}_s,\vec{\epsilon}_f,\tau)} + O(\Omega^{-1/2}),
\end{align}
where we have introduced the nondimensional ratio of slow and fast timescales
\begin{align}
\gamma=\frac{\tau_s}{\tau_f}.
\end{align}
Note that the order $\Omega^0$ is defined by the operators
\begin{subequations}
\begin{align} 
& \mathcal{L}_{s}= -\nabla_{s}^T \tilde{\matrix{J}}_{s} \vec{\epsilon}_{s} +  \frac{1}{2}\nabla_{s}^T \tilde{\matrix{D}}_{s} \nabla_{s}\,,\label{eqn:operatorDecomposition:Ls}\\
& \mathcal{L}_{f}= -\nabla_{f}^T \tilde{\matrix{J}}_{f} \vec{\epsilon}_{f} +  \frac{1}{2}\nabla_{f}^T \tilde{\matrix{D}}_{f} \nabla_{f}\,,\label{eqn:operatorDecomposition:Lf}\\
& \mathcal{L}_\text{int} = -\nabla_{{f}}^T \tilde{\matrix{J}}_{fs} \vec{\epsilon}_{s} + \frac{1}{2} \nabla_{{s}}^T \tilde{\matrix{D}}_{sf} \nabla_{{f}} \notag\\&\qquad-\nabla_{{s}}^T \tilde{\matrix{J}}_{sf} \vec{\epsilon}_{f} + \frac{1}{2} \nabla_{{f}}^T \tilde{\matrix{D}}_{fs} \nabla_{{s}}\,,\label{eqn:operatorDecomposition:Lint}
\end{align}\label{eqn:operatorDecomposition}
\end{subequations}
\noindent where $\tilde{\matrix{J}}_{s}=\matrix{S}_s(\nabla_{\vec{x}_s} \vec{f}^T)^T$, $\tilde{\matrix{J}}_{f}=\matrix{S}_f(\nabla_{\vec{x}_f} \vec{f}^T)^T$ and $\tilde{\matrix{J}}_{sf}=\matrix{S}_s(\nabla_{\vec{x}_f} \vec{f}^T)^T$, $\tilde{\matrix{J}}_{fs}=\matrix{S}_f(\nabla_{\vec{x}_s} \vec{f}^T)^T$ as well as $\tilde{\matrix{D}}_{s}=\matrix{S}_s\text{diag}(\vec{f})\matrix{S}_s^T$, $\tilde{\matrix{D}}_{f}=\matrix{S}_f\text{diag}(\vec{f})\matrix{S}_f^T$ and $\tilde{\matrix{D}}_{sf}=\tilde{\matrix{D}}_{fs}^T=\matrix{S}_s\text{diag}(\vec{f})\matrix{S}_f^T$.

\section{Derivation of the slow-scale linear noise approximation} \label{sec:ssLNA}

\subsection{Deterministic QSSA} \label{determ-QSSA}

The leading order terms, $O(\Omega^{1/2})$, of Eq. (\ref{eqn:cmeexp}) describe the dynamics of macroscopic concentrations and is given by the coupled set of {REs}:
\begin{subequations}
\begin{align}
 \frac{d\vec{x}_s}{d\tau}&=\matrix{S}_s \vec{f}(\vec{x}_s, \vec{x}_f)\,, \label{deg-eqs} \\
 \frac{1}{\gamma}\frac{d\vec{x}_f}{d\tau}&=\matrix{S}_f \vec{f}(\vec{x}_s, \vec{x}_f)\,. \label{adj-eqs}
\end{align} \label{REs}
\end{subequations}
The presence of timescale separation is reflected by the large parameter $\gamma$ diminishing the time derivative in Eq.~\reff{adj-eqs}. Such a set of equations present a special case in singular perturbation theory, where Eqs.~\reff{deg-eqs} and \reff{adj-eqs} for the slow and fast variables are typically referred to as the degenerate and adjoined systems, respectively \cite{heinrich-schuster-book-96}. Tikhonov's first theorem  \cite{tikhonov-52, klonowski-83} states that a simplification of the above equations under timescale separation conditions is possible whenever certain requirements are met: i) the solutions of both the degenerate and adjoined systems, Eqs.~\reff{REs} are unique and their right-hand sides are continuous functions; ii) the root $\vec{x}_f=h(\vec{x}_s,\tau)$ is the stable solution of the adjoined system; iii) the initial values $\vec{x}_f(\tau=0)$ are in the domain of influence of the solution as in ii). Whenever these prerequisites are met, the solution of the full system \reff{REs} 
for $\vec{x}_s$ tends to the solution of the reduced system
\begin{align}
& \frac{d\vec{x}_s}{d\tau}=\matrix{S}_s \vec{f}(\vec{x}_s, h(\vec{x}_s))\,, \label{reduced-REs}
\end{align}
in the limit of timescale separation, i.e., $\gamma^{-1} \rightarrow 0$. Note that $\vec{x}_f=h(\vec{x}_s)$ is the solution of $\matrix{S}_f \vec{f} (\vec{x}_s, \vec{x}_f)=0$. 

These requirements are typically fulfilled for the biochemical networks of interest. This is since the chemical transformation scheme \reff{reaction} is formulated for elementary reactions, which are bimolecular or simpler, the right hand sides of Eqs.~\reff{REs} are continuous polynomial functions of the second  order at most. For monostable networks, the rate equations admit a single steady state which is the same for the full and the reduced REs, i.e., Eq.~\reff{REs} and ~\reff{reduced-REs}. It is therefore clear that all the solutions will tend to this state with time, quicker for fast variables and slower for the slow ones. 

\subsection{Adiabatic elimination of stochastic variables using the projection operator formalism}

In the previous subsection, we reviewed how {imposing timescale separation on the deterministic level} leads to reduced time evolution equations for the concentrations of the slow species. A related question which is the main {issue} of this article is: what is the reduced Fokker-Planck equation describing the fluctuations about the concentrations predicted by the deterministic QSSA? 

The Fokker-Planck equation describing the fluctuations in the concentrations of both fast and slow species is given by the $O(\Omega^0)$ terms in Eq. (\ref{eqn:cmeexp}); this has the form

{\begin{align}
& \frac{\partial}{\partial\tau} \Pi(\vec\epsilon_s, \vec\epsilon_f, \tau) =\left(\gamma {\mathcal{L}}_f+{\gamma}^{1/2}{\mathcal{L}}_\text{int}+{\mathcal{L}}_s\right)\Pi(\vec\epsilon_s, \vec\epsilon_f, \tau)\,. \label{gard-L}
\end{align}}
From the definitions of the operators Eq. (\ref{eqn:operatorDecomposition}), it can be seen that ${\mathcal{L}}_f$ acts only on the fast variables, ${\mathcal{L}}_s$ acts only on the slow variables and ${\mathcal{L}}_{\rm{int}}$ acts only on both slow and fast variables. Hence the three operators describe processes evolving on fast, slow, and intermediate timescales, respectively. The dimensionless parameter $\gamma$ here weights the degree of timescale separation in the system. 
{The fact that $\gamma$ is the same as for the deterministic QSSA implies that the conditions for timescale separation of the stochastic variables,  in the limit of large $\Omega$, are exactly the same as those for the REs.}

For well-separated timescales, i.e., the case $\gamma\gg 1$, we are typically interested in the probability distribution of slow variables, $\Pi(\vec\epsilon_s,\tau)=\int d \vec\epsilon_f \Pi(\vec\epsilon_s, \vec\epsilon_f, \tau)$. Projection operator methods have been found useful in facilitating the adiabatic elimination of fast variables from stochastic descriptions \cite{zwanzig1960,gardiner-book-04}. Here we use one such method to rigorously obtain a reduced Fokker-Planck equation for the fluctuations in the concentrations of the slow variables. 

The main idea behind the projection operator method is that one specifies the quasi steady-state probability distribution $\pi(\vec\epsilon_f)$ of fast fluctuations, which is determined by the infinite $\gamma$-limit of Eq. (\ref{gard-L}): 
\begin{align}
\mathcal{L}_f \pi(\vec\epsilon_f) = 0. 
\end{align}
The {reduction is then carried out defining} the operator 
\begin{equation}
 \label{POPintform}
 (\mathcal{P} \Pi)(\vec\epsilon_s, \vec\epsilon_f, \tau)=\pi(\vec{\epsilon}_f)\int d\vec{\epsilon}_f\, \Pi(\vec\epsilon_s, \vec\epsilon_f, \tau)=\pi(\vec\epsilon_f) \Pi(\vec\epsilon_s, \tau),
\end{equation}
projecting the probability distribution $\Pi(\vec\epsilon_s,\vec\epsilon_f, \tau)$ onto the distribution of fast fluctuations evaluated at steady-state. Note that the above definition satisfies the relation $\mathcal{P}^2 = \mathcal{P}$ and hence $\mathcal{P}$ is indeed a projector.

In what follows we use the forms we have derived for the operators in the Fokker-Planck Eq. (\ref{gard-L}), i.e., those given by Eq. (\ref{eqn:operatorDecomposition}), to deduce three properties of the projection operator. Given these properties we then show how the projection operator applied to Eq. (\ref{gard-L}) leads to a reduced Fokker-Planck equation in the slow variables only.

\subsubsection*{Properties of the projection operator}

In this subsection we will show {that} the following properties hold
\begin{subequations}
\begin{align}
& \mathcal{P} {\mathcal{L}}_s   = {\mathcal{L}}_s \mathcal{P}\,,\label{gard-pc1}\\
& \mathcal{P} {\mathcal{L}}_f   = {\mathcal{L}}_f \mathcal{P}=0\,,\label{gard-pc2}\\
& \mathcal{P} {\mathcal{L}}_\text{int}\, \mathcal{P} = 0 \label{gard-pc3}\,.
\end{align}\label{gard-pcs}
\end{subequations}

First, property \reff{gard-pc1} follows from the fact that the projection operator $\mathcal{P}$, as defined by Eq.~\reff{POPintform}, acts only on the fast variables $\vec{\epsilon}_f$, whereas ${\mathcal{L}}_s$, see Eq.~(\ref{eqn:operatorDecomposition:Ls}), acts only on the slow variables $\vec{\epsilon}_s$ and hence the two operators $\mathcal{P}$ and $\mathcal{L}_s$ commute.

Second, we show that both equalities of property \reff{gard-pc2} are satisfied. Considering the left hand side, we obtain
\begin{align}
 \mathcal{P}{\mathcal{L}}_f=\pi(\vec{\epsilon}_f)\int d \vec{\epsilon}_f {\nabla_f}^T \left( ... \right)=0\,,
\end{align}
since ${\mathcal{L}}_f$, as given by Eq.~\reff{eqn:operatorDecomposition:Lf}, has the form of {a divergence in the fast variables} and hence by the partial integration lemma, its integral vanishes in the absence of boundary terms. By considering the right hand side, we have
\begin{align}
 {\mathcal{L}}_f \mathcal{P}=\left({\mathcal{L}}_f \pi(\vec{\epsilon}_f)\right)\int d \vec{\epsilon}_f =0\,,
\end{align}
\noindent by the quasi-steady state condition, ${\mathcal{L}}_f \pi=0$.

The third property \reff{gard-pc3} can be {obtained} as follows. {The first, second and fourth terms of ${\mathcal{L}}_\text{int}$} as given by Eq. (\ref{eqn:operatorDecomposition:Lint}) have the form of a divergence in the fast variables and hence by the partial integration lemma they give no contribution to $\mathcal{P} {\mathcal{L}}_\text{int} \mathcal{P}$. The third term of Eq. (\ref{eqn:operatorDecomposition:Lint}) also does not contribute albeit for a different reason than for the three other terms just {discussed}
\begin{align}
\mathcal{P} {\mathcal{L}}_\text{int} \mathcal{P}&=\pi(\vec{\epsilon}_f)\int d \vec{\epsilon}_f \left(
 -{\nabla}_{s}^T \tilde{\matrix{J}}_{sf} \vec{\epsilon}_{f}
\right) \pi (\vec{\epsilon}_f) \int d \vec{\epsilon}_f^{\,\,}\notag\\
&=-\pi(\vec{\epsilon}_f)
 {\nabla}_{s}^T \tilde{\matrix{J}}_{sf} \langle \vec{\epsilon}_f\rangle_\pi \int d \vec{\epsilon}_f^{\,\,}=0\,.
\end{align}
\noindent Here, we have applied the partial integration lemma and then used the fact that $\langle \vec{\epsilon}_f\rangle_\pi=0$. The latter follows from the explicit form of ${\mathcal{L}}_f$, see Eq.~\reff{eqn:operatorDecomposition}, which implies that the solution of the Fokker-Planck equation for the fast fluctuations ${\mathcal{L}}_f \pi(\vec{\epsilon}_f) = 0$ is a Gaussian distribution centered about zero. 

\subsubsection*{Derivation of the projection operator method}

We will now use the three properties of the projection operator just derived to obtain a reduced Fokker-Planck equation. Our approach in this subsection follows that of Gardiner \cite{gardiner-84,gardiner-book-04}.

We define $\mathcal{Q}=1-\mathcal{P}$ and the following two functions
\begin{align}
\label{eqn:fproj}
      v(\tau) \equiv \mathcal{P} \Pi(\vec\epsilon_s, \vec\epsilon_f, \tau),
\ \   w(\tau) \equiv \mathcal{Q}\Pi(\vec\epsilon_s, \vec\epsilon_f, \tau),
\end{align}
together with their Laplace transforms
\begin{align}
\label{eqn:laplacet} 
     \tilde{v}(s)=\int_0^\infty d \tau ~ e^{-s \tau} v(\tau), \ \
     \tilde{w}(s)=\int_0^\infty d \tau ~ e^{-s \tau} w(\tau).
\end{align}
The latter has the distinct advantage that instead of dealing with differential equations we obtain a set  {of algebraic equations}.
Using Eqs. (\ref{eqn:fproj}), (\ref{eqn:laplacet}) together with Eq. (\ref{gard-L}) we find
\begin{subequations}
\begin{align}
 s \tilde{v}(s) - v(0) &= \mathcal{L}_s \tilde{v}(s) + \gamma^{1/2}\mathcal{P} \mathcal{L}_\text{int} \tilde{w}(s), \\
 s \tilde{w}(s) - w(0) &= (\mathcal{L}_s + \gamma\mathcal{L}_f + \gamma^{1/2}\mathcal{Q}  \mathcal{L}_\text{int}) \tilde{w}(s) \nonumber \\ &+ \gamma^{1/2}\mathcal{Q} \mathcal{L}_\text{int} \tilde{v}(s).
\end{align}
\end{subequations}
Note that use has been made of the properties (\ref{gard-pc1}) and (\ref{gard-pc2}).
Solving for $\tilde{v}(s)$, we obtain:
\begin{align}
\label{eqn:vresult}
 s \tilde{v}(s) &-v(0){-}\gamma^{1/2} \mathcal{P}  \mathcal{L}_\text{int} \mathcal{D}(\gamma)w(0) \notag \\
&=\left[ \mathcal{L}_s + 
	\gamma \mathcal{P}  \mathcal{L}_\text{int} \mathcal{D}(\gamma) \mathcal{L}_\text{int}
  \right] \tilde{v}(s),
\end{align}
where we have used definition (\ref{eqn:fproj}) and property (\ref{gard-pc3}) and introduced $\mathcal{D}(\gamma)=(s-\mathcal{L}_s-\gamma\mathcal{L}_f-\gamma^{1/2}\mathcal{Q}  \mathcal{L}_\text{int})^{-1}$. From the above equation one can draw the limit $\gamma\to\infty$ for which $\mathcal{D}(\gamma)\sim -(\gamma \mathcal{L}_f)^{-1}$:
\begin{align}
 \frac{\partial}{\partial \tau} {v}(\tau) =\left[ \mathcal{L}_s - \mathcal{P}  \mathcal{L}_\text{int}\mathcal{L}_f^{-1}\mathcal{L}_\text{int}\right]{v}(\tau),
\end{align}
where we have inverted the Laplace transform Eq. (\ref{eqn:laplacet}).
Note that due to the vanishing of the third term on the left hand side of Eq. (\ref{eqn:vresult}) this asymptotic limit is Markovian and hence does not require the knowledge of the initial distribution $w(0)$ of the fast fluctuations. Using $v(\tau)=\pi(\vec{\epsilon_f})\Pi(\vec{\epsilon_s},\tau)$ and integrating over the fast fluctuations $\vec{\epsilon}_f$ we obtain
\begin{subequations}
\begin{align}
 &\frac{\partial \Pi(\vec\epsilon_s,\tau)}{\partial\tau}=\mathcal{L}'\Pi(\vec\epsilon_s,\tau)\,,\label{gard-FPE-red}\\
 &\mathcal{L}'={\mathcal{L}}_s-\langle {\mathcal{L}}_\text{int} \,{\mathcal{L}}_f^{-1}\,{\mathcal{L}}_\text{int}\rangle_\pi\,,
\label{gard-L-red}
\end{align} \label{gard-red}
\end{subequations}
where the angled brackets with subscript $\pi$ in Eq.~\reff{gard-L-red} denote the statistical average over the steady-state probability distribution $\pi(\vec\epsilon_f)$ of fast fluctuations. 

\subsubsection*{Derivation of the slow-scale linear noise approximation}

The above equation is a generic form for the Fokker-Planck equation for the slow fluctuations $\vec{\epsilon}_s$ under timescale separation conditions. What remains is to explicitly evaluate the average over $\pi(\vec\epsilon_f)$ such that we obtain a closed-form expression for the reduced Fokker-Planck equation. We now show these evaluation steps in detail. 

Using $\mathcal{L}_\text{int}$ as given by Eq.
\reff{eqn:operatorDecomposition} together with Eq. \reff{gard-L-red}
we can deduce the form of the reduced Fokker-Planck operator
\begin{align}
\mathcal{L}'=
	{\mathcal{L}}_s
&- 	\nabla_s^T \tilde{\matrix{J}}_{sf} \langle \vec\epsilon_f
{\mathcal{L}}_f^{-1} \nabla_f^{\,\,T} \rangle_\pi\,
\tilde{\matrix{J}}_{fs}\vec{\epsilon}_s \notag\\
&-	\nabla_s^T \tilde{\matrix{J}}_{sf} \langle \vec\epsilon_f
{\mathcal{L}}_f^{-1} \vec\epsilon_f^{\,\,T} \rangle_\pi\,
\tilde{\matrix{J}}_{sf}^T\nabla_s \notag\\
&+	\frac{1}{2}\nabla_s^T \tilde{\matrix{J}}_{sf} \langle
\vec\epsilon_f {\mathcal{L}}_f^{-1} \nabla_f^{\,\,T} \rangle_\pi\,
\tilde{\matrix{D}}_{fs}\nabla_s \notag\\
&+	\frac{1}{2}\nabla_s^T \tilde{\matrix{D}}_{sf} \langle
\vec\epsilon_f {\mathcal{L}}_f^{-1} \nabla_f^{\,\,T} \rangle_\pi^T\,
\tilde{\matrix{J}}_{sf}^T \nabla_s .
\label{an_red_FP}
\end{align}
Note that terms which have $\nabla_f^T$ to the left do not contribute to the reduced operator and hence are missing from the above equation. 

We proceed by evaluating the two distinct correlators appearing in the
above expression explicitly. We shall make use of the identity
\begin{align}
\int_0^{\infty} du \,e^{{\mathcal{L}}_f u}=\left.
{\mathcal{L}}_f^{-1}e^{{\mathcal{L}}_f
u}\right|_0^\infty=-{\mathcal{L}}_f^{-1}(1-\mathcal{P})\,,
\end{align}
\noindent which can be verified from straightforward integration and
the fact that $\mathcal{P}\Pi(\vec{\epsilon_s},\vec{\epsilon_f},\tau)=\lim_{u\to\infty}
e^{\mathcal{L}_f u} \Pi(\vec{\epsilon_s},\vec{\epsilon_f},\tau)$
\cite{gardiner-book-04}.
Using the fact that $\mathcal{P} \vec{\epsilon}_f^{\,\,T}\pi=0$, we can write
\begin{align}
\langle \vec{\epsilon}_f\, {\mathcal{L}}_f^{-1}\,
\vec{\epsilon}_f^{\,\,T}\rangle_\pi
&=\int d \vec{\epsilon}_f \, \vec{\epsilon}_f
\mathcal{L}_f^{-1}(1-\mathcal{P}) \vec{\epsilon}_f^{\,\,T}\pi\notag\\
&=-\int_0^\infty du\,\int d \vec{\epsilon}_f \, \vec{\epsilon}_f\,
e^{{\mathcal{L}}_f u}\,\vec{\epsilon}_f^{\,\,T}\,\pi\notag\\
&=-\int_0^\infty du\langle
\vec{\epsilon}_f(u)\vec{\epsilon}_f^{\,\,T}(0) \rangle_\pi.
\end{align}
Note that in the third step we have taken into account that
$\vec{\epsilon}_f^{\,\,T}\pi(\vec{\epsilon}_f,u)=e^{{\mathcal{L}}_fu}\vec{\epsilon}_f^{\,\,T}\pi(\vec{\epsilon}_f,0)$
is a solution to $\partial_u\pi={\mathcal{L}}_f\pi$ with the initial
condition $\vec{\epsilon}_f^{\,\,T}\pi(0)$.
One can utilize the Fourier transform of the autocorrelation matrix
$\langle \vec{\epsilon}_f(u) \vec{\epsilon}_f^{\,\,T}(0)
\rangle_\pi=\int d\omega/(2\pi)e^{i\omega u}\matrix{S}_f(\omega)$ to
find that
\begin{align}
\label{eqn:correlator1}
\langle \vec{\epsilon}_f\,
{\mathcal{L}}_f^{-1}\,\vec{\epsilon}_f^{\,\,T}\rangle_\pi=-\frac{1}{2}\matrix{S}_f(0)=-\frac{1}{2}\tilde{\matrix{J}}_f^{-1}\tilde{\matrix{D}}_f\tilde{\matrix{J}}_f^{-T}.
\end{align}
Similarly, one can show that
\begin{align}
\label{eqn:correlator2}
\langle \vec{\epsilon}_f\, {\mathcal{L}}_f^{-1}\,{\nabla}_f^T
\rangle_\pi 
& = -\int_0^{\infty}du \,e^{{\matrix{J}}_fu}\langle \vec{\epsilon}_f
{\nabla}_f^T\rangle_\pi=-\tilde{\matrix{J}}_f^{-1}\,.
\end{align}
\noindent Here, $\langle \vec{\epsilon}_f
{\nabla}_f^T\rangle_\pi=-\matrix{I}$ with identity matrix $\matrix{I}$
is evaluated by partial integration. Plugging Eqs.
(\ref{eqn:correlator1}) and (\ref{eqn:correlator2}) into Eq.
(\ref{an_red_FP}) gives a closed-form expression for the Fokker-Planck
operator of the reduced probability distribution
$\Pi(\epsilon_s,\tau)$:
\begin{subequations}
\begin{align}
\mathcal{L}'=& -\nabla_{s}^T \tilde{\matrix{J}} \vec\epsilon_{s} +
\frac{1}{2}\nabla_{s}^T \matrix{\tilde{D}}_{ss} \nabla_{s},\\
\tilde{\matrix{J}}=&\tilde{\matrix{J}}_s-\tilde{\matrix{J}}_{sf}\tilde{\matrix{J}}_f^{-1}\tilde{\matrix{J}}_{fs},\\
{\matrix{\tilde{D}}}_{ss}=&\tilde{\matrix{D}}_s+\tilde{\matrix{J}}_{sf}\left(\tilde{\matrix{J}}_f^{-1}\tilde{\matrix{D}}_f\tilde{\matrix{J}}_f^{-T}\right)\tilde{\matrix{J}}_{sf}^T\notag\\
&-\tilde{\matrix{J}}_{sf} \tilde{\matrix{J}}_f^{-1} \tilde{\matrix{D}}_{fs}
-(\tilde{\matrix{J}}_{sf} \tilde{\matrix{J}}_f^{-1} \tilde{\matrix{D}}_{fs})^T.
\end{align}
\end{subequations}
%

We complete our analysis by changing to natural variables of {concentration
fluctuations $\hat{\vec{\eta}}_{s}=(\tau_s/\Omega)^{1/2}\vec\epsilon_{s}$, of real time $t = \tau \tau_s$} and concentrations $\vec{\phi}_{s,f}=\tau_{s,f}\vec{x}_{s,f}$ to finally yield the reduced Fokker-Planck equation for the fluctuations in the slow species

{\begin{subequations}
\begin{align}
\label{eqn:app_ssLNA}
\frac{\partial}{\partial t}P(\hat{\vec{\eta}}_s,t)&=
\left(-\hat{\nabla}_{s}^T {\matrix{J}} \hat{\vec{\eta}}_{s} +
\frac{1}{2}\hat{\nabla}_{s}^T \matrix{D}_{ss}
\hat{\nabla}_{s}\right)P(\hat{\vec{\eta}}_s,t)\,,\\
\matrix{J}=&\matrix{J}_s-\matrix{J}_{sf}\matrix{J}_f^{-1}\matrix{J}_{fs}\,,
\label{redJ_ssLNA_appA}\\
\matrix{D}_{ss}=&\matrix{D}_s+\matrix{J}_{sf}\left(\matrix{J}_f^{-1}\matrix{D}_f\matrix{J}_f^{-T}\right)\matrix{J}_{sf}^T\notag\\
&-\matrix{J}_{sf} \matrix{J}_f^{-1} \matrix{D}_{fs}
-(\matrix{J}_{sf} \matrix{J}_f^{-1} \matrix{D}_{fs})^T\, \label{diff_ours},
\end{align}
\end{subequations}}
where $\hat{\nabla}_s$ denotes the derivative with respect to
$\hat{\vec{\eta}}_{s}$.
The coefficient matrices in the above expressions are given by
$\matrix{J}_{s}=\matrix{S}_s(\nabla_{\vec{\phi}_s} \vec{f}^T)^T$,
${\matrix{J}}_{f}=\matrix{S}_f(\nabla_{\vec{\phi}_f} \vec{f}^T)^T$ and
${\matrix{J}}_{sf}=\matrix{S}_s(\nabla_{\vec{\phi}_f} \vec{f}^T)^T$,
${\matrix{J}}_{fs}=\matrix{S}_f(\nabla_{\vec{\phi}_s} \vec{f}^T)^T$ as
well as ${\matrix{D}}_{s}=\Omega^{-1}\matrix{S}_s\matrix{F}\matrix{S}_s^T$,
${\matrix{D}}_{f}=\Omega^{-1}\matrix{S}_f\matrix{F}\matrix{S}_f^T$,
${\matrix{D}}_{sf}={\matrix{D}}_{fs}^T=\Omega^{-1}\matrix{S}_s\matrix{F}\matrix{S}_f^T$
and $\matrix{F}=\text{diag}(\vec{f})$.

We note that the slow-scale Jacobian \reff{redJ_ssLNA_appA} coincides
with the reduced Jacobian as obtained from the macroscopic QSSA
equations, i.e., Eqs. \reff{reduced-REs} as shown in Ref.
\cite{thomas-bmc-12}. It is also important to note that the reduced
diffusion matrix $\matrix{D}_{ss}$ admits the representation
\begin{equation}
\matrix{D}_{ss}= \Omega^{-1} (\matrix{A}-\matrix{B})(\matrix{A}
-\matrix{B})^T, \label{newD_appA}
\end{equation}
\noindent where $\matrix{A} = \matrix{S}_s \sqrt{\matrix{F}}$ and
$\matrix{B} = \matrix{J}_{sf} {\matrix{J}}_f^{-1} \matrix{S}_f
\sqrt{\matrix{F}}$. From this representation it can be immediately deduced 
that the reduced matrix $\matrix{D}_{ss}$ is symmetric and positive
semi-definite, two crucial properties of the diffusion matrices for
all {Fokker-Planck equations} \cite{vankampen-book-07}. 
Using standard methods \cite{gardiner-book-04} one can also obtain the Langevin equations
corresponding to the slow-scale {Fokker-Planck equation \reff{eqn:app_ssLNA}. These are given by}
\begin{equation}
\frac{d}{d t} \hat{\vec{\eta}}_s(t) = \matrix{J} \hat{\vec{\eta}}_s(t)
+ \Omega^{-1/2}
(\matrix{S}_s-\matrix{J}_{sf}\matrix{J}_{f}^{-1}\matrix{S}_f)
\sqrt{\matrix{F}} \vec{\Gamma}(t)\,, \label{LEssLNA_appA}
\end{equation}
\noindent where the $R$ dimensional vector $\vec{\Gamma}(t)$ is white
Gaussian noise defined by $\langle \Gamma_i(t) \rangle = 0$ and
$\langle \Gamma_i (t)\Gamma_j (t') \rangle = \delta_{i,j} \delta(t -
t')$. Equations (\ref{eqn:app_ssLNA}) and (\ref{LEssLNA_appA})
constitute equivalent forms of the LNA under timescale separation conditions, with the latter being particularly useful for Monte Carlo simulation purposes. This reduced LNA is precisely the ssLNA introduced {in Ref.} \cite{thomas-bmc-12}.

\section{Reaction networks with slow and fast reactions}  \label{sec:rel}

{Besides the QSSA there {exists} another popular method to eliminate the fast variables from the deterministic REs; this is the rapid-equilibrium approximation \cite{segel-93,heinrich-schuster-book-96,klipp-05}}. While the QSSA considers the situation in which there exists slow and fast species whose transients decay on well-separated timescales, the rapid-equilibrium approximation divides the set of reactions into groups of slow and fast reactions where the latter determines the equilibrium of the fast species alone \cite{heinrich-schuster-book-96}.

In the stochastic case there exist a variety of methods separating fast and slow reactions. Examples are the ``nested stochastic simulation algorithm'' \cite{weinan-07}, ``slow-scale stochastic simulation algorithm'' \cite{cao-etal-05}, the ``stochastic partial equilibrium assumption'' \cite{cao2-etal-05} and the ``quasiequilibrium approximation'' \cite{goutsias-05}. {Confusingly, also the ``stochastic quasi-steady state assumption'' of Rao and Arkin \cite{rao-03} is valid only in the limit of slow and fast reactions \cite{thomas-jcp-comm-11,sanft-11}.} All of these approaches have in common that the microscopic rate functions, or equivalently the {propensities are rearranged into two groups: those associated with $R_s$ slow reactions which occur rarely over a long period of time and those associated with $(R-R_s)$ fast reactions which occur frequently over a short period of time.} Then we can define a constant $\mu$ such that 
\begin{align}
 \label{eqn:propSeparation}
 \mu = \frac{\max(\hat{f}_1,\hat{f}_2, ... , \hat{f}_{R_s})}{\min(\hat{f}_{R_s+1},\hat{f}_{R_s+2}, ... , \hat{f}_{R}) }  \gg 1,
\end{align}
holds. For the sake of this article we will refer to approximations utilizing the above criterion as ``rapid equilibrium approximations''. Since {our present derivation of the ssLNA} is based only on the assumption of the presence of slow and fast species it can be used to investigate the latter approximation as a partial case. 

Using the size parameter $\mu$ we can write the vector of macroscopic rate functions as
\begin{align}
 \vec{f}(\vec{x}_s,\vec{x}_f)=\left(\vec{f}_s(\vec{x}_s,\vec{x}_f),\mu\vec{f}_f(\vec{x}_s,\vec{x}_f)\right),
\end{align}
where $\vec{f}_s(\vec{x}_s,\vec{x}_f)=(f_1,f_2, ... , f_{R_s})$ are the rates of the slow reactions and $\vec{f}_f(\vec{x}_s,\vec{x}_f)=\mu^{-1}(f_{R_s+1},f_{R_s+2}, ... , f_{R})$ are the rates of the fast reactions rescaled by the size parameter $\mu$. Note that here $\mu$ is determined by the infinite volume limit of Eq. \reff{eqn:propSeparation}. We can now partition the stoichiometric matrix into block matrices 
\begin{align}
 \matrix{S}=
\left[
\begin{array}{cc}
 \matrix{S}_{s} \\
 \matrix{S}_{f}
\end{array}
\right]
=
\left[
\begin{array}{cc}
 \matrix{S}_{s}^{(s)} & \matrix{S}_{s}^{(f)} \\
 \matrix{S}_{f}^{(s)} & \matrix{S}_{f}^{(f)}
\end{array}
\right],
\end{align}
discriminating {slow and fast reactions (superscript) as well as slow and fast species (subscript). The matrices $\matrix{S}_{s}^{(s)}$ and $\matrix{S}_{f}^{(f)}$ denote the stoichiometries of the slow and fast species in the slow and fast reactions, respectively, while $\matrix{S}_{s}^{(f)}$ and $\matrix{S}_{f}^{(s)}$ represent the stoichiometry of slow species in fast reactions and the stoichiometry of fast species in slow reactions.}

\subsection{Deterministic rapid-equilibrium approximation}

The macroscopic elimination starts from the {conventional REs}
\begin{subequations}
\begin{align}
 \frac{d \vec{x}_s}{d\tau} = \mu \matrix{S}_{s}^{(f)} \vec{f}_f + \matrix{S}_{s}^{(s)} \vec{f}_s,   \\
 \frac{1}{\gamma}\frac{d \vec{x}_f}{d\tau} = \mu \matrix{S}_{f}^{(f)} \vec{f}_f  +  \matrix{S}_{f}^{(s)} \vec{f}_s, 
\end{align}
\end{subequations}
which are similar to Eqs. (\ref{REs}) but {discriminate} slow and fast reactions by the size parameter $\mu$. It is clear that $\mu$ and $\gamma$ must be {of the same order} since slow and fast timescales are determined by the size of the rate functions. 

Further, we observe that for $\mu \to \infty$ the above equations are not immediately {of the form required by Tikhonov's theorem (compare Eqs. (\ref{REs}))}, and hence the adiabatic elimination is not immediately applicable unless we impose $\matrix{S}_{s}^{(f)}=0$. This condition implies that the populations of slow species are not changed in fast reactions and is also imposed throughout the literature in reducing stochastic slow-fast reaction networks \cite{cao-etal-05,cao2-etal-05,goutsias-05,pahlajani-etal-11}. Setting the time-derivative of the second equation to zero we can solve $\matrix{S}_{f}^{(f)} \vec{f}_f\approx0$ for $\vec{x}_f=h(\vec{x}_s) + O(\mu^{-1})$ to obtain the reduced system 
\begin{align}
 \label{eqn:ODEreactionreduced}
 \frac{d \vec{x}_s}{d t} = \matrix{S}_s^{(s)} \vec{f}_s(\vec{x}_s,h(\vec{x}_s)).
\end{align}
In the case where the equilibrium of the fast reactions $\matrix{S}_{f}^{(f)} \vec{f}_f=0$ is detailed balanced, the above approximation is called the deterministic rapid-equilibrium approximation \cite{heinrich-schuster-book-96}, i.e., the case when the fast reactions are given by a set of reversible reactions for which the forward and backward rates of each reaction cancel {each other}.

\subsection{Stochastic rapid-equilibrium approximation}

We can now apply the ssLNA to obtain the contribution of the fluctuations using the same conditions as used above for the deterministic rapid-equilibrium approximation. First, we make use of the condition $\matrix{S}_s^{(f)}=0$ to obtain the coefficients $\matrix{J}_{s}=\matrix{S}_s^{(s)}(\nabla_{\vec{\phi}_s} \vec{f}_s^T)^T$, ${\matrix{J}}_{f}=\mu\matrix{S}_f^{(f)}(\nabla_{\vec{\phi}_f} \vec{f}_f^T)^T+\matrix{S}_f^{(s)}(\nabla_{\vec{\phi}_f} \vec{f}_s^T)^T$ and ${\matrix{J}}_{sf}=\matrix{S}_s^{(s)}(\nabla_{\vec{\phi}_f} \vec{f}_s^T)^T$, ${\matrix{J}}_{fs}=\matrix{S}_f^{(s)}(\nabla_{\vec{\phi}_s} \vec{f}_s^T)^T+\mu\matrix{S}_f^{(f)}(\nabla_{\vec{\phi}_s} \vec{f}_f^T)^T$ which distinguish contributions from slow and fast reactions. Second, we can use these together within Eq. (\ref{redJ_ssLNA_appA}) to obtain the reduced Jacobian by taking the limit $\mu \to \infty$
\begin{align}
 \matrix{J}&=\matrix{S}_s^{(s)}(\nabla_{\vec{\phi}_s} \vec{f}_s^T)^T - \notag\\
 &\matrix{S}_s^{(s)}(\nabla_{\vec{\phi}_f} \vec{f}_s^T)^T
 \left[\matrix{S}_f^{(f)}(\nabla_{\vec{\phi}_f}\vec{f}_f^T)^T \right]^{-1}
 \matrix{S}_f^{(f)}(\nabla_{\vec{\phi}_s} \vec{f}_f^T)^T.
\end{align}
It can be shown that the above expression coincides with the {Jacobian obtained} using Eq. (\ref{eqn:ODEreactionreduced}).
Third, we calculate the coefficients of the noise from Eq. \reff{newD_appA} as
\begin{align}
 \matrix{A}=\matrix{S}_s^{(s)}\sqrt{\matrix{F}_s}, \ \
 \matrix{B}=0,
\end{align}
where $\matrix{F}_s=\text{diag}(\vec{f}_s)$ and $\matrix{F}_f=\text{diag}(\vec{f}_f)$. Note that the second equation follows from inserting $\matrix{S}_f\sqrt{\matrix{F}}=\matrix{S}_f^{(s)}\sqrt{\matrix{F}_s}+\mu \matrix{S}_f^{(f)}\sqrt{\matrix{F}_f}$ into the definition of $\matrix{B}$ after Eq. (\ref{newD_appA}) together with the expression for ${\matrix{J}}_{f}$ and ${\matrix{J}}_{sf}$ and taking the limit $\mu \to \infty$. This implies that $\matrix{B}$ is of order $\mu^{-1/2}$ and hence the noise stemming from the fast reactions can be neglected in the limit $\mu\to\infty$.
Finally, we can formulate the Langevin equations
\begin{equation}
\frac{d}{d t} \hat{\vec{\eta}}_s(t) = \matrix{J} \hat{\vec{\eta}}_s(t) + \Omega^{-1/2}
\matrix{S}_s^{(s)}\sqrt{\matrix{F}_s} \vec{\Gamma}(t)\,. \label{LEreactionreduced}
\end{equation}
Note that these are consistent with those obtained using stoichiometry $\matrix{S}_s^{(s)}$ and propensity vector $\vec{f}_s(\vec{x}_s,h(\vec{x}_s))$ of the reduced macroscopic Eqs. \reff{eqn:ODEreactionreduced}. {Note that while in the ssLNA, Eq. \reff{LEssLNA_appA}, (which is consistent with QSSA conditions) both the noise in the fast and slow reactions contribute to the noise of the slow species, in the stochastic rapid-equilibrium approximation, Eq. \reff{LEreactionreduced}, the noise in the slow reactions alone determines that in the slow species.} 

{The latter Langevin equation has also been obtained by Pahlajani et al. \cite{pahlajani-etal-11} starting from the decomposition of reactions into slow and fast categories. However our derivation is the first to clearly show that this Langevin equation is a partial case of the ssLNA and hence is only valid over a subset of the parameter space over which the QSSA holds.}

\section{Discussion}

In this article we have shown how to rigorously reduce the linear noise approximation of the CME by using the projection operator formalism to eliminate the fast fluctuation variables. 
The resulting Langevin equation, Eq.\reff{LEssLNA_appA}, is in agreement with the ssLNA as has been previously derived from intuitive arguments only \cite{thomas-bmc-12}. The present derivation provides a rigorous basis by deriving the LNA from the system size expansion using a modified van Kampen ansatz, Eq. \reff{LNA-ansatz} which is applicable under conditions of timescale separation.
The resulting REs, Eq. \reff{REs} and the Fokker-Planck {equation} \reff{gard-L} obtained by this approach are of a particular form which allows direct application of Tikhonov's theorem and the projection operator method, respectively. Hence by this procedure it is guaranteed that the mesoscopic elimination of the fast fluctuations is valid in exactly the same limit as the macroscopic elimination of the concentrations of the fast species by the deterministic QSSA.

For reaction networks composed of slow and fast reactions, conditions considered by the majority of available stochastic model reduction methods employing timescale separation, we have shown that the limit of large volumes 
or molecule numbers the CME can be approximated by a Langevin equation, Eq. \reff{LEreactionreduced}, which is a special case of the ssLNA. The main advantage of using the ssLNA over the various aforementioned methods \cite{weinan-07,cao-etal-05,cao2-etal-05,goutsias-05,rao-03,pahlajani-etal-11} is that the ssLNA is valid over a larger parameter range than the latter methods. The path integral approach developed in \cite{gts-11} also enjoys this property. However the ssLNA enjoys the further advantage that it is available in closed form for any monostable reaction network and hence can be readily constructed from knowledge of the stoichiometric matrix and deterministic rate equations. The disadvantage of the ssLNA is that for pathways composed of some second order reactions, the ssLNA (and the LNA on which it is based) is valid only for large enough molecule numbers \cite{grima-09,thomas-12}. This limitation can be lifted by consideration of higher order terms in the {system size expansion}; 
such calculations present more formidable analytical challenges than encountered in the derivation of the ssLNA and are under current investigation. 

{
For realistic biochemical networks there may be particular parameter ranges for which the stability of the dynamics is either monostable or bistable or even oscillatory states can occur. The applicability of the method therefore depends on the type of stability realized for a particular network under consideration. Oscillatory states are found for $10$\% of the transcriptome and $20$\% of the proteome in mouse liver \cite{panda-02,reddy-06} and similar fractions in the human metabolome \cite{dallmann-12}. Although bimodal distributions have been observed experimentally, as for instance in the \emph{lac} operon of E. coli \cite{choi-08}, a recent proteome-wide study suggests that such probability distributions (which potentially indicate bistability) are quite rare \cite{taniguchi-10} and similarly for the human transcriptome \cite{mason-11}. 
Despite the fact that bistable and oscillatory properties are important for specific cellular functions, it appears that monostable networks for which the present theory has been developed are common in living cells.
}

{
We note that the LNA has been applied also to networks with limit cycles \cite{scott-06}. The resulting equation is still a linear Fokker-Planck equation and hence the elimination of fast variables can be performed along the same lines as in the present derivation. However, the analysis of the resulting equation has to be carried out by means of Floquet theory due to the inherent phase diffusion in these systems \cite{boland-09}. %
For bistable systems, the underlying distribution cannot be captured by a linear Fokker-Planck Eq. and hence the LNA is not applicable in this case \cite{vankampen-book-07}. A commonly employed procedure to eliminate fast variables in such systems is the stochastic QSSA. A recent numerical study reported considerable discrepancies in the probability distributions of full and QSSA-reduced bistable systems \cite{agarwal-12}. We expect these discrepancies to be similar to the difference in the ssLNA and rapid-equilibrium approximations discussed in this article for the monostable case. Therefore, developing a technique to rigorously reduce the CME of bistable networks remains still an open question.}
{
Since our method is devised for monostable systems and can be extended to oscillatory ones under timescale separation conditions we expect it to be of broad applicability for the study of intracellular reaction networks.
}

\section*{Acknowledgments}

R. G. acknowledges support by SULSA (Scottish Universities Life Science Alliance).


\begin{thebibliography}{10}

\bibitem{alon-book-07}
U.~Alon,
\newblock {\em An introduction to systems biology: design principles of
  biological circuits} (CRC Press, London, 2007).

\bibitem{alon-nature-07}
U.~Alon,
\newblock Nature {\bf 446}, 497 (2007).

\bibitem{segel-slemrod-89}
L.A.~Segel and M.~Slemrod,
\newblock SIAM Rev. {\bf 31}, 446 (1989).

\bibitem{lee-othmer-09}
C.H.~Lee and H.G.~Othmer,
\newblock J. Math. Biol. {\bf 60}, 387 (2010).

\bibitem{grima-08}
R.~Grima and S.~Schnell,
\newblock Essays Biochem. {\bf 45}, 41 (2008).

\bibitem{janssen-89}
J.A.M.~Janssen,
\newblock J. Stat. Phys. {\bf 57}, 187 (1989).

\bibitem{mastny-07}
E.A.~Mastny, E.L.~Haseltine, and J.B.~Rawlings,
\newblock J. Chem. Phys. {\bf 127}, 094106 (2007).

\bibitem{thomas-jcp-comm-11}
P.~Thomas, A.V.~Straube, and R.~Grima,
\newblock J. Chem. Phys. {\bf 135}, 181103 (2011).

\bibitem{sanft-11}
K.R.~Sanft, D.T.~Gillespie, and L.~Petzold,
\newblock IET Syst. Biol. {\bf 5}, 58 (2011).

\bibitem{thomas-bmc-12}
P.~Thomas, A.V.~Straube, and R.~Grima,
\newblock BMC Syst. Biol. {\bf 6}, 39 (2012).

\bibitem{heinrich-schuster-book-96}
R.~Heinrich and S.~Schuster,
\newblock {\em The regulation of cellular systems} (Chapman {\&} Hall, New
  York, 1996).

\bibitem{vankampen-book-07}
N.G.~van Kampen,
\newblock {\em Stochastic processes in physics and chemistry} (Elsevier,
  Amsterdam, 2007).

\bibitem{gillespie-rev-07}
D.T.~Gillespie,
\newblock Annu. Rev. Phys. Chem. {\bf 58}, 35 (2007).

\bibitem{gillespie-92}
D.T.~Gillespie,
\newblock Physica A {\bf 188}, 404 (1992).

\bibitem{gillespie-09}
D.T.~Gillespie,
\newblock J. Chem. Phys. {\bf 131}, 164109 (2009).

\bibitem{tikhonov-52}
A.N.~Tikhonov,
\newblock Mat. Sb. {\bf 31}, 575 (1952),
\newblock (in Russian).

\bibitem{klonowski-83}
W.~Klonowski,
\newblock Biophys. Chem. {\bf 18}, 73 (1983).

\bibitem{zwanzig1960}
R.~Zwanzig,
\newblock J. Chem. Phys. {\bf 33}, 1338 (1960).

\bibitem{gardiner-book-04}
C.W.~Gardiner,
\newblock {\em Stochastic methods: a handbook for the natural and social
  sciences} (Springer, Berlin, 2010).

\bibitem{gardiner-84}
C.W.~Gardiner,
\newblock Phys. Rev. A {\bf 29}, 2814 (1984).

\bibitem{segel-93}
I.H.~Segel,
\newblock {\em Enzyme kinetics: behavior and analysis of rapid equilibrium and
  steady state enzyme systems} (Wiley, Hoboken, 1993).

\bibitem{klipp-05}
E.~Klipp,
\newblock {\em Systems biology in practice: concepts, implementation and
  application} (Wiley, Hoboken, 2005).

\bibitem{weinan-07}
E.~Weinan, D.~Liu, and E.~Vanden-Eijnden,
\newblock J. Comp. Phys. {\bf 221}, 158 (2007).

\bibitem{cao-etal-05}
Y.~Cao, D.T.~Gillespie, and L.R.~Petzold,
\newblock J. Chem. Phys. {\bf 122}, 014116 (2005).

\bibitem{cao2-etal-05}
Y.~Cao, D.T.~Gillespie, and L.R.~Petzold,
\newblock J. Comp. Phys. {\bf 206}, 395 (2005).

\bibitem{goutsias-05}
J.~Goutsias,
\newblock J. Chem. Phys. {\bf 122}, 184102 (2005).

\bibitem{rao-03}
C.V.~Rao and A.P.~Arkin,
\newblock J. Chem. Phys. {\bf 118}, 4999 (2003).

\bibitem{pahlajani-etal-11}
C.D.~Pahlajani, P.J.~Atzberger, and M.~Khammash,
\newblock J. Theor. Biol. {\bf 272}, 96 (2011).

\bibitem{gts-11}
N.A.~Sinitsyn, N.~Hengartner, and I.~Nemenman,
\newblock Proc. Natl. Acad. Sci. {\bf 106}, 10546 (2009).

\bibitem{grima-09}
R.~Grima,
\newblock BMC Syst. Biol. {\bf 3}, 101 (2009).

\bibitem{thomas-12}
P.~Thomas, H.~Matuschek, and R.~Grima,
\newblock PLoS ONE {\bf 7}, e38518 (2012).

\bibitem{panda-02}
S.~Panda et al.,
\newblock Cell. {\bf 109}, 307--320 (2002).

\bibitem{reddy-06}
A.B.~Reddy et al.,
\newblock Curr. Biol. {\bf 16}, 1107--1115 (2006).

\bibitem{dallmann-12}
R.~Dallmann et al.,
\newblock Proc. Natl. Acad. Sci. {\bf 109}, 2625--2629 (2012).

\bibitem{choi-08}
P.J.~Choi, L.~Cai, K.~Frieda, and X.S.~Xie,
\newblock Science {\bf 322}, 442--446 (2008).

\bibitem{taniguchi-10}
Y.~Taniguchi et al.,
\newblock Science {\bf 329}, 533--538 (2010).

\bibitem{mason-11}
C.C.~Mason et al.,
\newblock BMC Genomics {\bf 12}, 98 (2011).

\bibitem{scott-06}
M. Scott, B. Ingalls, and M. K\ae{}rn,
\newblock Chaos {\bf 16}, 026107 (2006).

\bibitem{boland-09}
R.P.~Boland, T.~Galla, and A.J.~McKane,
\newblock Phys. Rev. E {\bf 79}, 051131 (2009).

\bibitem{agarwal-12}
A.~Agarwal, R.~Adams, G.C.~Castellani, and H.Z.~Shouval,
\newblock J. Chem. Phys. {\bf 137}, 044105 (2012).

\end{thebibliography}
\end{document}